\documentclass[twocolumn,secnumarabic,nobibnotes,prd,superscriptaddress,nofootinbib]{revtex4-2}

\usepackage{amsmath,amsfonts,amssymb,bm}
\usepackage[scr=rsfs]{mathalpha}

\newcommand{\beq}{\begin{equation}} \newcommand{\eeq}{\end{equation}}
\newcommand{\bea}{\begin{eqnarray}} \newcommand{\eea}{\end{eqnarray}}
\newcommand{\lb}{\label}

\usepackage[T1]{fontenc}
\usepackage[utf8]{inputenc}
\usepackage[english]{babel}
\usepackage[dvipsnames]{color}

\newcommand{\FURG}{\affiliation{ Instituto de Matemática, Estatística e Física, Universidade Federal do Rio Grande, Avenida Itália km 8, 96201-900, Rio Grande, Rio Grande do Sul, Brazil Brazil}}
\newcommand{\UFABC}{\affiliation{Centro de Ci\^encias Naturais e Humanas, Universidade Federal
do ABC,  Avenida dos Estados 5001, 09210-580 Santo Andr\'e, São Paulo, Brazil}}

\begin{document}

\title{Generalized nonconservative gravitational field equations from Herglotz action principle}

\author{Juilson A. P. Paiva}\UFABC

\author{Matheus J. Lazo}\FURG

\author {Vilson T. Zanchin}\UFABC

\begin{abstract}

We present an alternative nonconservative gravitational theory based on the Herglotz variational principle in a fully covariant form. The present model may be seen as an improvement of the theory proposed in Ref.~[Lazo et al, Phys. Rev. D {\bf 95}, 101501 (2017)], whose resulting theory is meaningful just in particular coordinate systems. In the present work, we report a new theory that is free from such a restriction. It is also obtained using the Herglotz variational principle and by taking advantage of the restricted equivalence between Lagrangian functions in the scope of such action principle. The more restricted class of equivalent Lagrangian functions, in comparison with the Hamilton variational principle, is the key point to find a Lagrangian that furnishes a new alternative gravitational theory that is fully covariant. Once the equations that govern the dynamics of the gravitational field are obtained, a few simple cosmological models are investigated. It is found that the Herglotz gravitational field reduces to a single function that, under certain conditions, plays the role of the cosmological constant in general relativity, turning unnecessary the use of dark energy to explain the accelerated expansion of the universe.
The linearized version of the theory is also investigated and it is verified that the theory shows a dissipative character in regard to gravitational waves. From observational data, in both scenarios, the Herglotz vector field is estimated.
\end{abstract}
\maketitle

\section{Introduction}

The principle of stationary action, or principle of least action, together with theorems relating symmetry properties of the action to conserved physical quantities of a given system, as the Noether theorem, form the basis of modern theoretical physics. Despite its significant contribution to the progress of theoretical physics, such a principle leaves behind some gaps regarding the description of all possible physical systems. An important gap is the lack of a general formulation for dissipative systems. Alternative formulations have been tried over the years with relative success, at least for some nonconservative mechanical systems \cite{Galley:2012hx} (see also, e.g., \cite{Chandra,Bender,Lin} and references therein). An interesting alternative is a formulation due to Herglotz \cite{Herg1,Herg2,GGB,MJJG2}, a variational principle that appropriately describes mechanical systems with damping forces. The Herglotz principle may also be extended to classical fields \cite{GGB,MJJG2,MJG} allowing to obtain, e.g., the electromagnetic field equations in dissipative media \cite{MJJG2}. 

Interestingly, in Ref.~\cite{Lazo:2017udy} a nonconservative gravity theory was proposed based on the Herglotz variational problem. However, the resulting field equations of this theory are not manifestly covariant in the sense that they involve some extra terms that depend on the frame's choice. In fact, the proposal theory introduces a new vector field $\lambda_\mu$ which couples to the action density field $s_\mu$. The noncovariance is a consequence of choosing a nonscalar Lagrangian for the geometry sector, in the tensorial sense, so that the additional terms in the field equations involving $\lambda_\mu$ are not tensorial functions. Despite that fact, recent results show that this nonconservative gravity is a promising alternative theory to dark energy \cite{Fabris:2017msx,Carames:2018atv,Fabris:2020wkr}.  In particular, Ref.~\cite{Fabris:2017msx} investigates the correspondence between a cosmological solution within such a modified gravity theory and the universe filled with a viscous cosmological fluid.

Since its recent formulation, the nonconservative gravity obtained through the Herglotz variational principle \cite{Lazo:2017udy} has been considered in several contexts \cite{Fabris:2017msx,Carames:2018atv,Fabris:2020wkr, Braganca:2018elt, Ayuso:2020vuu,Bravetti:2020tau,Olmo:2019flu,Fabris:2019qvy,Vermeeren:2019ovq,Faraoni:2018qdr, Fabris:2018nli,Daouda:2018kuo,Bravetti:2018rts}. Besides the applications to mimic the dark energy in cosmological models mentioned above \cite{Fabris:2017msx, Carames:2018atv,Fabris:2020wkr}, other interesting applications have been made, e.g., to build models for compact objects \cite{Fabris:2019qvy}, in braneworld gravity \cite{Fabris:2018nli}, and to model cosmic string configurations \cite{Braganca:2018elt}. In Ref.~\cite{Fabris:2019qvy} (see also \cite{Olmo:2019flu}), the authors analyze the existence of compact object solutions, a work that motivated investigations also on wormhole solutions~\cite{Ayuso:2020vuu}. Additionally, in Ref.~\cite{Faraoni:2018qdr} the possible correspondence between scalar-tensor gravity theories of Brans-Dicke type and Lagrangian descriptions of dissipation such as the gravity theory derived from the Herglotz principle \cite{Lazo:2017udy} was evidenced. These and other studies on this subject are now motivating further research on generalized nonconservative gravity theories.

Nonconservative gravity theories have been proposed over the years for a variety of reasons. For instance, Rastall \cite{rastall} argued that energy-momentum conservation could be a valid phenomenon only in flat spacetime, and proposed a modified theory of gravity that appears to be nonconservative. See, however, Ref.~\cite{lindblom} for a criticism of this and other similar theories, and Refs.~\cite{wolf,Visser:2017gpz} for further considerations and other references on this subject.  See also  Ref.~\cite{Velten:2021xxw} for a recent review on nonconservative gravity theories. 

 Pursuing the idea of formulating a consistent nonconservative gravity theory, we follow here a similar path as done in \cite{Lazo:2017udy}, but keeping control of the assumptions to obtain a theory whose field equations are formulated in a manifestly covariant form. Actually, in the present work, we show that the Herglotz action principle introduced in \cite{Lazo:2017udy} provides two direct possibilities to formulate nonconservative gravity theories. The first possibility is by considering only first-order derivatives in the Lagrangian function, as done in \cite{Lazo:2017udy} and that leads to a noncovariant and nonconservative theory of gravity. 
 The second possibility, that we consider in the present work, is to consider up to second-order derivatives in the Lagrangian. As we are going to show, this second approach yields a consistent and covariant nonconservative gravity theory.

The present work is structured as follows. In the next section, the Herglotz variational problem is presented and some of its features of interest for the present study are briefly discussed. Section~\ref{sec:garvityaction} is devoted to the formulation of a nonconservative gravity theory based on the Herglotz problem. In Section~\ref{sec:cosmology},  some different cosmological solutions are presented and analyzed, and an estimate for the Herglotz parameter is given. The linear approximation of the theory is obtained in Section~\ref{sec:linear} and the damping effects on gravitational waves are confirmed. Finally, in Sec.~\ref{sec:conclusion} we make further considerations about the properties of the formulated theory and conclude.

\section{An action principle for nonconservative systems}

\subsection{The Herglotz variational principle}
\label{sec:Herglotz}

In recent works, a physically meaningful action principle for nonconservative systems was proposed \cite{Lazo:2017udy,MJJG2}. It was first employed to obtain a nonconservative gravity theory \cite{Lazo:2017udy} and then it was extended to other fields and other kinds of systems \cite{MJJG2}. This action principle is obtained from a generalization of the Herglotz variational problem, introduced in 1930 by Herglotz \cite{Herg1,Herg2,GGB}, in order to include fields as a function of several independent variables. The basic idea of this generalized variational problem is to consider a Lagrangian function depending, itself, on the action.

The original formulation due to Herglotz \cite{Herg1} applies to the classical dynamics and consists in the problem of determining the function $x(t)$ that extremizes the functional $S(b)$, where the action $S(t)$, with $t\in [a,b]\in \mathbb{R}$, is a solution of the problem
\beq \lb{H}
\dot{S}(t)=L(t,x(t),\dot{x}(t),S(t)),
\eeq
under the boundary conditions
\beq\lb{boundH}
S(a)=S_a,\;x(a)=x_a,\;x(b)=x_b, 
\eeq
with the overdot standing for the total derivative with respect to the parameter $t$. 
It is important to stress that $S(t)$ is a functional since, for each function $x(t)$, it follows a different differential equation. Therefore, $S(t)$ depends on the function $x(t)$. Furthermore, the Herglotz variational problem \eqref{H}-\eqref{boundH} reduces to the classical fundamental problem of the calculus of variations when the Lagrangian function $L$ does not depend on $S(t)$. In this particular case, by integrating \eqref{H} it results in the classical variational problem, which consists of extremizing the functional
\beq
\lb{H2}
S(b)=\int_a^b \bar{L}(t,x(t),\dot{x}(t))\;dt,
\eeq
where $a<b$, $x(a)=x_a$, $x(b)=x_b$ are fixed endpoints, and
\beq
\lb{H3}
\bar{L}(t,x(t),\dot{x}(t))=L(t,x(t),\dot{x}(t)) +\frac{S_a}{b-a}.
\eeq
This is, of course, equivalent to the Hamilton variational principle.

Herglotz \cite{Herg1,Herg2} proved that a necessary condition for a function $x(t)$ to yield an extreme for the variational problem \eqref{H} is to satisfy the generalized Euler-Lagrange equation
\beq
\lb{HEL}
\frac{\partial L}{\partial x} -\frac{d}{dt}\frac{\partial L}{\partial \dot{x}}+\frac{\partial L}{\partial S}\frac{\partial L}{\partial \dot{x}}=0.
\eeq
It is clear that in the case where ${\partial L}/{\partial S}=0$, as in the classical problem \eqref{H2}, the differential equation \eqref{HEL} reduces to the usual Euler-Lagrange equation, which is obtained from the Hamilton principle.

The potential of the Herglotz problem for applications to nonconservative systems is evident even in the simplest case, where the dependence of the Lagrangian function on the action is linear \cite{MJJG2}. For instance, the Lagrangian functional
\beq
\lb{H3b}
L=\frac{m\dot{x}^ 2}{2}-U(x)-\frac{\gamma}{m} S
\eeq
describes a dissipative system of a pointlike particle of mass $m$ under the potential $U(x)$ and submitted to a viscous force proportional to the velocity. In fact, the resulting equation of motion that follows from Eqs. \eqref{H3b} and \eqref{HEL},
\beq
\lb{H3c}
m\ddot{x}+\gamma \dot{x}=F,
\eeq
 where $\ddot{x}$ is the particle acceleration and $F=-{d U}/{dx}$ is the external force, includes the well-known dissipative term proportional to the velocity $\dot{x}$, and whose resistance coefficient is $\gamma$. In this context, the linear term $-{\gamma} S/{m}$ in the Lagrangian \eqref{H3b} can be interpreted as a potential function for the nonconservative force, see \cite{MJJG2}. Furthermore, the Lagrangian given by \eqref{H3b} is physical in the sense that it provides us with physically meaningful relations for the momentum and for the Hamiltonian (see, e.g., \cite{MJJG2,Riewe,LazoCesar}).

The formulation of an action principle in terms of the Herglotz variational problem \eqref{H}-\eqref{boundH}, instead of the traditional calculus of variation problem \eqref{H2}, has two direct justifications. The first one is the fact that the Herlotz problem enables us to formulate a physically meaningful Lagrangian problem for nonconservative systems displaying linear first-order time derivative forces, like the frictional force in \eqref{H3c}. The second justification is the fact that, in any physical theory, the Lagrangian function which defines the action is constructed from the scalars (invariant quantities) of the theory. Consequently, since the action itself is a scalar, the most general Lagrangian may itself be a function of the action \cite{Lazo:2017udy}.

\subsection{Equivalence of Lagrangians according to the Herglotz principle}

In the classical Hamilton action principle, the Lagrangian describing a physical system is not uniquely defined. Two Lagrangian functions $L$ and $\tilde{L}$ are said to be equivalent if they establish the same Euler-Lagrange equations. However, in general, this equivalence does not hold in the context of the Herglotz action principle.

To verify this fact we consider the Herglotz problem with the Lagrangian $\tilde{L}=L+\dot{f}$. It thus consists in extremizing the functional $\tilde{S}(b)$, but now the action $\tilde{S}(t)$ is such that
\beq
\label{eqv5}
\dot{\tilde{S}}(t)=L(t,x(t),\dot{x}(t),\tilde{S}(t))+\dot{f}(t,x(t)),
\eeq
subject to the same boundary conditions as in Eq.~\eqref{boundH}, with  
fixed $\tilde{S}(a)=\tilde{S}_a$. Notice that the function $f$  does not depend on the variables $\dot{x}$ and $\tilde S$, since the total Lagrangian for the standard Herglotz problem \eqref{H}  may depend only on $t$, $x(t)$, $\dot{x}(t)$, and $S(t)$.

Now, from \eqref{HEL} the Herglotz problem with the Lagrangian function $\tilde{L}=L+\dot{f}$ yields the following Euler-Lagrange equation, 
\begin{equation}
\label{eqv8}
\dfrac{\partial L}{\partial x}-\dfrac{d}{dt}\left(\dfrac{\partial L}{\partial \dot{x}}\right)+\dfrac{\partial L}{\partial \dot{x}}\dfrac{\partial L}{\partial S}\dfrac{\partial S}{\partial\tilde{S}}
+\frac{\partial f}{\partial x}\frac{\partial L}{\partial  \tilde{S}}=0,
\end{equation}
where we have used the identities
\beq 
\frac{d}{dt}\frac{\partial \dot{f}}{\partial \dot{x}}=\frac{d}{dt}\frac{\partial } {\partial \dot{x}}\left[\frac{\partial f}{\partial t}+\frac{\partial f}{\partial x}\dot{x}\right]=\frac{d}{dt}\frac{\partial f}{\partial x}=\frac{\partial \dot{f}}{\partial x}\, .
\eeq
Equation \eqref{eqv8}, in general, is different from the Euler-Lagrange equation for the Herglotz problem with Lagrangian $L$, Eq.~\eqref{HEL},  unless
$\partial L/\partial\tilde S=0$, where we recovered the classical variational problem, or the conditions \begin{equation}
  \frac{\partial f}{\partial x}=0, \quad \dfrac{\partial S}{\partial\tilde{S}}=1, 
\end{equation}
satisfy simultaneously.

\section{Action and equations of motion of a generalized nonconservative gravity}
\label{sec:garvityaction}

Let us start this section by reviewing the equivalence between Lagrangian functions in the context of the classical (Hamilton) action principle for the general relativity theory. As well known, in such a context there are two mostly used equivalent Lagrangian densities for gravity. The first is the Einstein--Hilbert Lagrangian $\mathcal{L}_g$ given by 
\beq \lb{H8b}
\begin{split}
& \mathcal{L}_g(x^\mu,g_{\alpha \beta},g_{\alpha \beta,\mu},g_{\alpha \beta,\mu,\nu}) = g^{\mu\nu}R_{\mu\nu} \equiv \tilde{\cal L}-{\cal L}_{ef}
\\ &\quad =g^{\mu\nu}(\Gamma^{\sigma}_{\mu\sigma,\nu}-\Gamma^{\sigma}_{\mu\nu,\sigma})-g^{\mu\nu}(\Gamma^{\sigma}_{\mu\nu}\Gamma^{\rho}_{\sigma\rho}-\Gamma^{\rho}_{\mu\sigma}\Gamma^{\sigma}_{\nu\rho}),
\end{split}
\eeq
where $g^{\mu\nu}$ is the inverse metric tensor, $R_{\mu\nu}$ is the Ricci tensor,  and we defined $\tilde{\cal L}=g^{\mu\nu}(\Gamma^{\sigma}_{\mu\sigma,\nu}-\Gamma^{\sigma}_{\mu\nu,\sigma})$ and ${\cal L}_{ef}=g^{\mu\nu}(\Gamma^{\sigma}_{\mu\nu}\Gamma^{\rho}_{\sigma\rho}-\Gamma^{\rho}_{\mu\sigma}\Gamma^{\sigma}_{\nu\rho})$.
The second commonly used Lagrangian is ${\cal L}$ itself.
The equivalence between these two Lagrangian functions is verified by noting that it holds the relation $\mathcal{L}_g=2{\cal L}_{ef}+\nabla_\mu J^\mu$,  where $J^\mu =g^{\mu\nu}\Gamma_{\nu\sigma}^\sigma-g^{\sigma\nu}\Gamma_{\sigma\nu}^\mu$. Consequently, $\mathcal{L}_g$ and $2\mathcal{L}$ differ from a divergence term, and by integrating over a given subset  $\cal {V}$ of the $n-$dimensional spacetime manifold $\cal {M}$ it gives, $\int_{\cal {V}}{\cal L}_g \sqrt{-g} \, d^{n}x=2\int_{\cal {V}}{\cal  L}_{ef} \sqrt{-g}\, d^{n}x$ plus a surface term (see, e.g., \cite{Dirac}), where $g$ is the determinant of the metric, demonstrating that ${\cal  L}_g$ and $2{\cal  L}_{ef}$ are equivalent Lagrangian functions according to the Hamilton variational principle.

Now, since Lagrangian functionals differing from a total derivative are not equivalent in the Herglotz variational problem (and consequently are not equivalent when differing from a divergence term), we have two simple possibilities to formulate a dissipative gravitational theory in such a context.
The first possibility, investigated in Ref.~\cite{Lazo:2017udy}, is by taking $\mathcal{L}_{ef}$ as the gravitational part of the Lagrangian.
The second possibility, that we explore in the present work, is by choosing $\mathcal{L}_{g}$, with $\mathcal{L}_{g}$ given in Eq.~\eqref{H8b}.

The approach considered in \cite{Lazo:2017udy} has the interesting mathematical advantage of the Lagrangian being function depending only on first-order derivatives of the metric tensor. It considers a Lagrangian function given by $\mathcal{L}=\mathcal{L}_{ef}+\lambda_\nu s^\nu$, 
 where $\lambda_\mu=\lambda_\mu(x)$ is an arbitrary vector field\footnote{\!We name it as the Herglotz vector field, or the Herglotz parameter.}, and $s^\mu$ is the action density vector field (see \cite{Lazo:2017udy,MJJG2}).  However, this approach has the physical disadvantage that the Lagrangian $\mathcal{L}_{ef}$ is not a scalar density and, consequently, the resulting field equations for the theory are valid only in a specific set o referential frames fixed a priory.

Inspired by Ref.~\cite{Lazo:2017udy}, here we consider an alternative proposal by taking the Lagrangian as $\mathcal{L}=\mathcal{L}_{g}+\lambda_\nu s^\nu$.  Since $\mathcal{L}_{g}$ is an invariant (scalar density), the field equations of our theory will be given by truly tensorial equations and it is not necessary to fix a preferential coordinate system a priory as in \cite{Lazo:2017udy}. However, since $\mathcal{L}_{g}$ has second-order derivatives of the metric tensor it is necessary to impose additional boundary conditions on the metric in order to fix the variational problem. The derivation of the field equations from such a Lagrangian is presented next. 

Let the spacetime be defined as an n-dimensional smooth manifold ${\cal M}$ endowed with a Lorentzian metric $g_{\mu\nu}$. Now let $\mathcal{V}$ be a subset of ${\cal M} $ with boundary $\Omega$, which is considered as a Jordan surface whose unit normal vector is denoted by $n_\mu$. Then, the generalized action principle may be stated in terms of the functional $S(\Omega)$ given by (see also \cite{Lazo:2017udy}),
\beq
\label{eq1}
\begin{split}
S(\Omega)&=\int_{\Omega}n_\mu s^\mu \sqrt{|h|}\; d^{n-1}x=\int_{\mathcal 
V} {s^\mu}_{;\mu}\,d^{n}x,\\
{s^\mu}_{;\mu} &=\mathcal{L}(x^\nu,g_{\alpha \beta},g_{\alpha \beta,\nu},s^\nu), 
\end{split}
\eeq
where $s^\mu$ is a differentiable action-density vector field,  the semicolon (;) stands for a covariant derivative, and $h$ is the determinant of the induced metric on $\Omega$. The boundary conditions we impose, in order to close the variational problem, is by keeping both the metric $g_{\mu \nu}$ and its derivatives $g_{\mu \nu,\gamma}$ fixed on $\Omega$.

We are going to consider a generalized Lagrangian given by $\mathcal{L}=F(x) \mathcal{L}_m+\mathcal{L}_{g}+\lambda_\nu s^\nu$, where ${\cal L}_m$ stands for the standard matter Lagrangian, and $F(x)$ is a coupling factor that may be a function of the coordinates. Hence, from \eqref{eq1} and \eqref{H8b} it follows that the action density $s^\mu$ is subjected to the additional condition
\begin{equation}
\left(s^\mu\sqrt{-g}\right)_{\!,\mu}=\sqrt{-g}\,\left( R_{\mu\nu}g^{\mu\nu}
+\lambda_\mu s^\mu+F\mathcal{L}_m\right)\label{lagran},
\end{equation}
where the comma stands for a partial derivative.

Our goal is to obtain the field equations for $g_{\mu\nu}$ whose solutions make the functional $S(\Omega)$ stationary under the condition \eqref{lagran}. Taking the variation of \eqref{eq1}  and \eqref{lagran} with respect to $g^{\mu\nu}$ it gives, respectively,
\begin{equation} \label{e2}
\begin{split} 
&\delta S(\Omega)=\int_{\Omega}n_\mu \delta\left(s^\mu\sqrt{h}\,\right)d^{n-1}x=0,\\
&{\zeta^\mu}_{,\mu}  
=\delta\left(R_{\mu\nu}g^{\mu\nu}\sqrt{-g}\,+ F\mathcal{L}_m\sqrt{-g}
\right) + \lambda_\mu\zeta^\mu,
\end{split}
\end{equation}
where $\zeta^\mu=\delta\left(s^\mu\sqrt{-g}\right)$. As considered in \cite{Lazo:2017udy}, since the hipersurface $\Omega$ and, consequently, $\sqrt{h}$ are fixed, i.e., they do not depend on the metric variation, we obtain from the first equation in \eqref{e2} that $\delta s^\mu =0$ on $\Omega$. On the other hand, the last equation in \eqref{e2} can be written as
\begin{equation} \label{e4}
\begin{split}
& \left(\zeta^\mu e^{-\phi}\right)_{\!,\mu}  = 
e^{-\phi}\delta\left(R_{\mu\nu}g^{\mu\nu}\sqrt{-g}+F\mathcal{L}_m\sqrt{-g}
\right) ,
\end{split}
\end{equation}
with $\phi=\int\lambda_\mu(x) dx^\mu$. 
Integrating the left hand side of \eqref{e4} over ${\cal V}$, and then working out the variation of $s^\mu\sqrt{g}$, it follows
\begin{equation}\label{e5}
\begin{split}
&
\int_{\cal V}
 \left(\zeta^\mu e^{-\phi}\right)_{\!,\mu}d^n x = \\ & 
\int_{\cal V}
\left[ \sqrt{-g}\,\left(\delta s^\mu -\dfrac{ s^\mu}{2} g_{\nu\sigma}\delta 
g^{\nu\sigma}\right) e^{-\phi}\right]_{\!,\mu}d^nx  = \\
&\int_{\Omega}n_\mu\left(\delta s^\mu -\dfrac{ s^\mu}{2}
g_{\nu\sigma}\delta 
g^{\nu\sigma}\right) e^{-\phi}\sqrt{h}\,d^{n-1} x=0, 
\end{split}
\end{equation}
where we have used the fact that $\delta s^\mu=0$ on $\Omega$, and we imposed the usual condition in the variational procedure that the metric field is fixed on the boundary $\Omega$, i.e., since $g_{\mu \nu}({\Omega})$ is fixed then $\delta g^{\nu\sigma}$ vanishes on $\Omega$. Consequently, the last integral in Eq.~\eqref{e5} gives zero.  Thus, taking these results back into Eq.~\eqref{e4}, after integration, we find 
\begin{equation}\begin{split}
\int_{\cal V} e^{-\phi}\left[g^{\mu\nu} \delta R_{\mu\nu}
+G_{\mu\nu}\delta 
g^{\mu\nu}\right]\sqrt{-g}d^nx+\\ \int_{\cal V} 
\, e^{-\phi} F\,\delta\left(\mathcal{L}_m\sqrt{-g}\right)d^n x=0,\label{e6}
\end{split}
\end{equation}
where $G_{\mu\nu}$ is the Einstein tensor. 

Let us now consider the first term in the last integral. Using the definition of the Ricci tensor in terms of the metric, we get (see, e.g.,  \cite{Carroll})
\begin{equation}
 \begin{split}
& \int_{\cal V} e^{-\phi}g^{\mu\nu}\delta R_{\mu\nu}\sqrt{-g}d^nx= \\ 
&  \int_{\cal V} 
e^{-\phi}\left[\left(g_{\mu\nu}
 g^{\sigma\gamma}\left(\delta g^{\mu\nu}\right)_{;\gamma} -\left(\delta 
g^{\sigma\gamma}\right)_{;\gamma}\right)\sqrt{-g}\right]_{,\sigma} d^n x =\\
 &\int_{\cal V} e^{-\phi}\left[\left(g_{\mu\nu}
 g^{\sigma\gamma}\left(\delta g^{\mu\nu}\right)_{;\gamma} -\left(\delta 
g^{\sigma\gamma}\right)_{;\gamma}\right)\sqrt{-g}\right]\lambda_{\sigma} 
d^nx, \label{eq7}
 \end{split}
 \end{equation}
 where an integration by parts was performed, and we consider the additional boundary condition imposed in the problem that $g_{\mu \nu,\gamma}({\Omega})$ is fixed (and consequently $(\delta g^{\mu\nu})_{;\gamma}$ vanishes on the boundary $\Omega$).

Let us now work out the terms on the right-hand side (RHS) of the last relation in Eq.~\eqref{eq7}. 
 The first integral term reads
 \begin{equation}
 \begin{split}
  &\int_{\cal V} e^{-\phi}g_{\mu\nu}
 g^{\sigma\gamma}\left(\delta g^{\mu\nu}\right)_{;\gamma}\lambda_{\sigma}\sqrt{-g} d^nx=\\
 &\int_{\cal V} e^{-\phi}g_{\mu\nu}
\lambda^\gamma\left(\delta g^{\mu\nu}_{,\gamma}+\Gamma^\nu_{\sigma\gamma}\delta g^{\mu\sigma}+\Gamma^\mu_{\sigma\gamma}\delta g^{\sigma\nu}\right)\sqrt{-g} d^nx= \\
  &\int_{\cal V} e^{-\phi}g_{\mu\nu}\left(\lambda^{\sigma}\lambda_{\sigma} -\lambda^\rho_{\;,\rho } +\Gamma^\sigma_{\gamma\rho}\lambda^{\rho} \right)\delta g^{\mu\nu}\sqrt{-g} d^nx, 
 \label{e8}
  \end{split}
 \end{equation}
 where an integration by parts was performed and the boundary terms were neglected once again.
 Applying the same procedure to the second term on the RHS of Eq. 
\eqref{eq7} one gets
 \begin{equation}
 \begin{split}
 &\int_{\cal V} e^{-\phi}\left(\delta 
g^{\sigma\gamma}\right)_{;\gamma}\lambda_{\sigma}\sqrt{-g}\, d^nx =  \\
&\int_{\cal V} e^{-\phi}\left(\delta 
g^{\sigma\gamma}_{,\gamma}+\Gamma^\gamma_{\mu\gamma}\delta 
g^{\sigma\mu}+\Gamma^\sigma_{\mu\gamma}\delta 
g^{\mu\gamma}\right)\lambda_{\sigma}\sqrt{-g}\, d^nx=\\
 &\int_{\cal V} e^{-\phi}\left(\lambda_{\mu}\lambda_{\nu}-\lambda_{\mu,\nu}+\Gamma^\sigma_{
\mu\nu}\lambda_{\sigma}\right)\delta g^{\mu\nu}\sqrt{-g}\, d^nx.\label{e9}
  \end{split}
 \end{equation}
 Now we put the results given by Eqs.~\eqref{e8} and \eqref{e9} back into \eqref{eq7} to obtain
 \begin{equation} \begin{split}
 & \int_{\cal V} e^{-\phi}\left(g^{\mu\nu}\delta R_{\mu\nu}- K_{\mu\nu}\delta g^{\mu\nu}\right)\sqrt{-g}\, d^n x=0 , \label{e10}
 \end{split}
 \end{equation}
where we introduced the tensor $K_{\mu\nu}$ given by 
\begin{equation} \label{eq:K}
K_{\mu\nu}=\Lambda_{\mu\nu}- g_{\mu\nu}\Lambda, 
\end{equation}
with $\Lambda_{\mu\nu}$ being the symmetric tensor
\begin{equation} \label{eq:Lambda}
\Lambda_{\mu\nu}
=\frac{1}{2}\left(\lambda_{\mu;\nu}+\lambda_{\nu;\mu}\right)-\lambda_{\mu}\lambda_{\nu},
\end{equation}
and $\Lambda= \Lambda_\mu^\mu$ is its trace.

Finally, from \eqref{e6} and 
\eqref{e10} we find the generalized field equations
\begin{equation}
R_{\mu\nu}-\frac{1}{2}g_{\mu\nu}R+K_{\mu\nu}=
\frac{F}{2}\, T_{\mu\nu},\label{EE}
\end{equation}
where  $F=F(x)$ is a non-negative arbitrary function that, in the conservative Einstein-Hilbert actions, plays the role of the (Newtonian) gravitational coupling constant, and the energy-momentum tensor is given by
\begin{equation}
T_{\mu\nu}=-\frac{2}{ \sqrt{-g}}\frac{\delta\left(\mathcal{L}_m\sqrt{-g}\right)}{\delta g^{\mu\nu}}.
\end{equation}
The usual Einstein field equations are recovered in the case $\lambda_\mu =0$, 
as long as we take $F= 16\pi G/c^4$, with $G$ being the universal 
gravitational constant.

Since $\lambda_\mu$ is a tensor (a vector) by definition, and the covariant derivative of a vector field is also a tensor, it becomes clear that $\Lambda_{\mu\nu}$ is a tensor. Thus, Eq.~\eqref{EE} is written in a manifestly covariant form.  

It is worth noticing that in an empty spacetime region, i.e., for $T_{\mu\nu}=0 $, or by imposing conservation of the matter energy-momentum tensor $\left(F\,T_{\mu}^\nu\right)_{;\nu}=0$, the Bianchi 
identity implies that the tensor $K_{\mu\nu}=\Lambda_{\mu\nu}-g_{\mu\nu}\Lambda$ also satisfies the  conservation condition $K_{\mu\;;\nu}^\nu=0$. However, the Herglotz principle is adapted to dissipative systems, for which the energy momentum tensor does not satisfy the conservation equation, and so the Bianchi identity gives  $\left(F\,T_{\mu}^\nu\right)_{;\nu}= 
{\Lambda_\mu^\nu}_{;\nu} - \delta_\mu^\nu \Lambda_{,\nu}$. Note that even by assuming energy-momentum conservation, tensor $K_{\mu\nu}$ may be a nonconserved quantity because of the presence of the function $F(x)$, i.e.,  $K_{\mu\;;\nu}^\nu = T_\mu^\nu F_{,\nu}$. This fact was explored in 
Ref.~\cite{Lazo:2017udy}, where the accelerating effect of dark energy in standard cosmological models was simulated by a time-dependent coupling function $F=F(t)$, $t$ being the cosmological time (see also Ref.~\cite{Fabris:2017msx}). In the present work we take $F$ as the usual coupling of general relativity, $F = 16\pi G/c^4$, but allow $G$ to be an arbitrary function of coordinates.

\section{Application to cosmology}
\label{sec:cosmology}

\subsection{The modified Friedmann equations}

In order to investigate the consequences of the Herglotz vector field $\lambda_{\mu}$, we analyse the dynamics of a cosmological model filled with a perfect fluid, whose the energy-momentum tensor is of the form
\begin{equation}\label{emt}
T_{\mu\nu}=\left(\rho+p\right)u_\mu u_\nu+p\,g_{\mu\nu},
\end{equation}
where $\rho$ is the energy density, $p$ is pressure, and $u^\mu$ is the four-velocity of the fluid. For now, we consider the Friedmann-Lamaître-Robertson-Walker (FLRW) metric with zero space curvature, which may be written as
\begin{equation}\label{metric-flrw}
ds^2=-dt^2+a^2(t)\left(dr^2 + r^2d\Omega^2\right) ,
\end{equation}
where $a(t)$ is the scale factor, $t$ being the comoving time coordinate, and $d\Omega^2=d\theta^2+\sin^2\theta d\varphi^2$ is the metric on the unit sphere.

Given that the metric \eqref{metric-flrw} represents a spatially homogeneous and isotropic space-time, it admits a set of Killing vectors that generate the isometries. These vectors are useful for fixing the general form of the  Herglotz vector field $\lambda_\mu$ in this spacetime. For that, the $\lambda_\mu$ vector needs to satisfy the Killing equation
\begin{equation}
    \mathcal{L}_\xi\lambda_\mu=0, \label{eq:killing}
\end{equation}
where $\xi$ stands for the set of Killing vectors of the metric \eqref{metric-flrw}. After a detailed analysis, we realize that the most general vector $\lambda_\mu$ that satisfies Eq.~\eqref{eq:killing} is of the form
\begin{equation}\label{lambda-phi}
    \lambda_\mu=\left(\phi(t),0,0,0\right),
\end{equation}
with $\phi$ being a smooth function of the time $t$ only.

Substituting the vector field~\eqref{lambda-phi}, the metric \eqref{metric-flrw}, and the energy-momentum tensor~\eqref{emt} into Eq.~\eqref{EE}, and using \eqref{eq:Lambda},  we get the modified Friedmann equations for the scale factor
 \begin{eqnarray}
& &3\left(\dfrac{\dot{a}}{a}\right)^2-3\dfrac{\dot{a}}{a}\phi={8\pi G}\rho            \label{eqd:density} , \\
& & 2\dfrac{\ddot{a}}{a}+\left(\dfrac{\dot{a}}{a}\right)^2-2\dfrac{\dot{a}}{a}\phi+\phi^2-\dot{\phi}=-8\pi G\, p.\quad \label{eqd:pressure}
\end{eqnarray}
This is a system of two equations for four unknown functions. Hence, even after establishing, as usual,  an equation of state for the cosmological fluid, additional conditions are needed.  
 
Using the Bianchi identities and proceeding with the idea that $G$ may not be a constant (see, e.g, \cite{Uzan:2002vq}) it follows just one non-trivial relation, namely,
\begin{equation}
    3\left(\frac{\ddot{a}}{a}+\frac{\dot{a}}{a}\phi\right)\phi+8\pi\dot{G}\rho=-8\pi G\left[ \dot{\rho}+3\frac{\dot{a}}{a}\left(p+\rho\right)\right].\label{eq:div}
\end{equation}
 The right-hand side of Eq.~\eqref{eq:div} represents the covariant divergence of the energy-moment tensor, coupled to gravity through the function $G$. Note that, as expected, the energy-momentum conservation may be violated in the present theory even in the case of constant $G$. By assuming energy-momentum conservation, it follows
\begin{equation}
    3\left(\frac{\ddot{a}}{a}+\frac{\dot{a}}{a}\phi\right)\phi+8\pi\dot{G}\rho=0.\label{eq:emc1}
\end{equation}

It is worth emphasizing that the function $\phi (t)$ is arbitrary and hence a new cosmological model is built for every choice of that function. In fact, this freedom may be somehow fixed by imposing some physical conditions required by the cosmological model under construction.
For instance, considering an expanding cosmological model ($\dot a/a>0)$, in order to guarantee the non-negativity of energy density, the constraint 
\begin{equation}
   \dfrac{\dot{a}}{a} -\phi\geq 0
\end{equation}
must be obeyed at least for sufficiently large times. In the following, we analyze some particular simple cosmological models emerging from the present theory.

 \subsection{A conservative cosmological model}
 \label{sec:inflat}

 Although the theory introduces a nonconservative geometric gravitational aspect, in this section we explore the existence of solutions that deviate from general relativity in cases where the energy-momentum tensor is conserved and the coupling strength $G$ is constant.
 
 Considering energy-momentum conservation and the constancy of $G$, equation \eqref{eq:emc1} results in
\begin{equation}
\phi=- \frac{\ddot{a}}{\dot{a}}.\label{eq:phi}
\end{equation}
 After introducing this result into \eqref{eqd:density}, the energy density reads
 \begin{equation}
   {8\pi G}\,\rho = 3\left(\dfrac{\dot{a}}{a}\right)^2+3\dfrac{\ddot{a}}{a}.\label{eq:density1} 
 \end{equation}
 Now, by analyzing Eq.~\eqref{eq:density1} one concludes that any solution with accelerated expansion ($\ddot a>0$) provides a positive definite energy density. 
 
 On the other hand, by introducing \eqref{eq:phi} in \eqref{eqd:pressure} the expression for the pressure results in the form
 \begin{equation}
  {8 \pi  G}\, p=  -\left(\frac{4 \ddot{a}}{a}+\frac{\dot a^2}{a^2}+\frac{a^{(3)}}{\dot a}\right),\label{eq:pressure1}
 \end{equation}
 where $a^{(3)}$ stands for the third-order derivative of the scale factor with respect to time $t$.

The system of equations to be solved is now formed by Eqs.~\eqref{eq:phi}, \eqref{eq:density1}, and \eqref{eq:pressure1}. There are three equations for four unknowns. The usual strategy is to pick up an equation of state for the cosmic fluid. However, in the present case, such a strategy leads to a nonlinear third-order differential equation for the scale factor which has no solution in closed form. Hence, for simplicity, and since it is not our objective in the present work to consider the most general solution for Eqs.~\eqref{eq:phi}, \eqref{eq:density1}, and \eqref{eq:pressure1}, we follow the simpler road of choosing the explicit form of one of the unknown functions.  The first choice is a power-law function for the scale factor. In this case, we get, 
\begin{align}
&    a(t)= a_0 \left(\dfrac{ t} {t_0}\right)^\alpha, \label{eq:scale}\\
& \phi (t)= \dfrac{1-\alpha}{t},\\
& 8\pi\, G\,\rho(t) = \dfrac{3 \alpha\left(2\alpha-1\right)}{t^2}, \label{eq:density2}\\
& 8\pi\, G\, p(t) = \dfrac{\alpha\left(7-6\alpha\right)-2}{t^2}, \label{eq:pressure2}
\end{align}
where $\alpha$ is a constant parameter. 

Assuming that the scale factor is increasing with time ($\alpha>0$), the non-negativity of the energy density $\rho$ implies the constraint $\alpha\geq 1/2$. The cosmic fluid is well defined for all values of $\alpha$ in the interval $1/2\leq \alpha <\infty$, with the ratio ${\cal R}=p(t)/\rho(t)$ being independent of time and varying with $\alpha$ from ${\cal R}=1/3$ (for $\alpha$ close to $1/2$) to ${\cal R}=-1$ (in the limit $\alpha\to \infty)$. 

For $\alpha$ smaller than $1/2$ the energy-density and the pressure assume only negative values. 

Taking $\alpha=1/2$, $a(t)= a_0 \left(t/t_0\right)^{1/2}$, the energy density and pressure vanish, $p(t)=\rho(t)=0$,  and $\phi$ reduces exactly to the Hubble function, that is, $\phi=\dot{a}/a=1/2t$. 
This case reproduces exactly the same behavior of the spatially flat FLRW model dominated by radiation in general relativity. 
Thus, in vacuum or when the field $\phi$ dominates, the Herglotz field $\phi$ behaves like a fictitious radiation component, thinking of $\phi$ as a fictitious source, it simulates a cosmological model in general relativity with a perfect fluid whose effective energy density and pressure obey the equation of state $p_{ef}=\rho_{ef}/3 \sim a^{-4}(t) \sim t^{-2}$.  

By taking $\alpha=2/3$ it follows  that  $a(t)= a_0 \left(t/t_0\right)^{{2}/{3}}$, and $p=0$. This solution is known as the Einstein-de Sitter universe with cold dark matter (CDM). 
The equivalent (effective) in general relativity is a cosmic fluid obeying the equation of state  $p_{ef}=0$. 

For $\alpha$ larger than $2/3$ the pressure assumes only is negative values.
 
 Then, under the assumption that the energy-momentum tensor is conserved and that $G$ is constant, we get cosmological models which are equivalent to general relativity models.
 
 In the case where $\phi$ is constant, Eq.~\eqref{eq:phi} can be integrated for the scale factor yielding
 \begin{equation}
     a(t)=a_0\left(e^{-\phi t}-1\right), \label{eq:scale1}
 \end{equation}
 where $a_0$ is the integration constant and the big bang was chosen at  $t=0$. It is clear that the constant $\phi$ must be negative ($\phi<0$) so that the scale factor is in accordance with the present observational data  
 
 For the scale factor \eqref{eq:scale1}, energy density and pressure are given, respectively, by   \begin{equation}
 \begin{split}
  8 \pi  G\, \rho&=  \frac{3 \phi ^2 \left(2-e^{t \phi }\right)}{ \left(1-e^{t \phi }\right)^2},\\
 8 \pi  G\,  p&=\frac{\phi ^2 \left(6 e^{t \phi }-e^{2 t \phi }-6\right)}{ \left(1-e^{t \phi }\right)^2}.\label{eq:denspre}
   \end{split}
 \end{equation}
 
Note that, although the energy density is non-negative, the pressure is always negative for any cosmological time. The ratio ${\cal R}(t) = p(t)/\rho(t)$ varies with time from ${\cal R}(t)=-1/3$ (for $t\to 0$) to ${\cal R}(t)=-1$ (for $t\to\infty$).
Therefore, applying this solution to the beginning of times, the result is an inflationary model governed by a fluid of cosmic strings $p\simeq -\rho/3$. On the other hand, applying the solution to very late times, the result is an accelerated expansion driven by a cosmological constant. In fact, for very large times, the components of the fluid \eqref{eq:denspre} reduce to  $8\pi G\,p=-8\pi G\, \rho=-6\,\phi ^2$, which is the same equation of state for a fluid represented by the cosmological constant $\Lambda$ in general relativity, with $\Lambda=6\,\phi^2$. 

The asymptotic limit of solution \eqref{eq:scale1} allows us to estimate the value of the Herglotz field $\phi$ at present epoch. In fact, in a FLRW model (within general relativity) dominated by the cosmological constant one has $H_0^2=\Lambda/3$. Hence, considering that the present solution gives $\phi^2=\Lambda/6$ we get $|\phi|=\sqrt{2}H_0/2\simeq 0.71\, H_0$. Then, using the value of $H_0$ obtained, for instance, from Ref.~\cite{Hotokezaka:2018dfi}, $H_0\sim 70$ km/s/Mpc, it follows $|\phi| \simeq  50$ km/s/Mpc $\simeq 1.6 \times 10^{-18}{\rm s^-1}$.

\subsection{A nonconservative model: An accelerated expanding phase dominated by cold dark matter}
\label{sec:CDM}

Sticking to the case of constant $G$, here we investigate the possibility of building models for accelerated expansion within the present theory without recurring to the mysterious dark energy content.
To take the simplest road, we put the pressure to zero, $p=0$, so it results in a cold dark matter dominated phase. Even after substituting the ansatz \eqref{eq:dSphase} into Eqs.~\eqref{eqd:density} and \eqref{eqd:pressure}, one degree of freedom is available. Again aiming at a simple model, let us assume that, during a given phase of the cosmic expansion, the scale factor $a(t)$ may be approximate by a growing exponential function, 
\begin{equation}\label{eq:dSphase}
    a(t) = a_0 e^{ht}, 
\end{equation}
with $a_0$ and $h$ being constant parameters, and with $h>0$.
 After this choice,  Eq.~\eqref{eqd:pressure} furnishes,
\begin{equation}\label{eq:phi1}
    \phi(t) = h  +\sqrt{2}\, h \tan\left[\sqrt{2}\, h\left(\, t - t_1\right)\right],
\end{equation}
where $t_1$ is an integration constant. Substituting \eqref{eq:phi1} into \eqref{eqd:density} it follows,
\begin{equation}\label{rho1}
    \rho(t) = \frac{3\sqrt{2}}{8\pi G}\,h^2  \tan\left[\sqrt{2}\, h\left(\, t_1 - t\right)\right],
\end{equation}
while the pressure is zero. 
This phase of accelerated expansion is generated by a CDM model.

Since $t_1$ is an arbitrary integration constant, its value may be adjusted so that the accelerated expansion phase lasts long enough to conform the present   observational data. However, to guarantee the non-negativity of the energy density, in the present case, the accelerated expansion cannot last forever after. There must be a mechanism to turn on the field $\phi(t)$  at the time $t\equiv t_0=t_1-\pi /2\sqrt{2}h$, and to turn it off just before the time $t =t_1$.
Once this mechanism is activated, its duration is at most a time interval given by $\Delta t=\pi/(2\sqrt{2}\, h)$ until it is turned off. It is still necessary to adjust the constant $t_1$ in favor of explaining the accelerated expansion at the current cosmological time.

\section{Linear approximation}
\label{sec:linear}

\subsection{Linearized theory}

Here we consider the metric resulting from a small perturbation around the Minkowski spacetime, i.e.
\begin{equation}
\begin{split}
g_{\mu\nu}&=\eta_{\mu\nu}+h_{\mu\nu}, \quad \|h_{\mu\nu}\| \ll 1,
\end{split}
\end{equation}
were $\eta_{\mu\nu}$ is the Minkowski metric tensor, and the quantities $h_{\mu\nu}$ are perturbation functions. We also assume that the Herglotz vector field is perturbed around its background value $\bar\lambda_\mu$, which is the solution of Eq.~\eqref{EE} in Minkowski spacetime, i.e.,
\begin{equation}
\begin{split}
\lambda_\mu&=\bar \lambda_\mu+l_\mu, 
\end{split}
\end{equation}
where $l_\mu$ is a perturbation on the background Herglotz vector $\bar \lambda_\mu$. 

Therefore, up to first order in the perturbations $h_{\mu\nu}$ and $l_\mu$, the tensor $K_{\mu\nu}$ defined in Eq.~\eqref{eq:K} may be split as a background and a perturbation term,
\begin{equation}
    K_{\mu\nu}= \bar{K}_{\mu\nu} + k_{\mu\nu}, \label{eq:K01}
\end{equation} 
where $ \bar{K}_{\mu\nu}$ and $ k_{\mu\nu}$ are given respectively by
\begin{equation} \label{eq:K01b}
    \begin{split}
\bar K_{\mu\nu} = &\; \bar \Lambda_{\mu\nu}- \eta_{\mu\nu}\bar\Lambda, \\
 k_{\mu\nu} = &\; \ell_{\mu\nu} - \eta_{\mu\nu}\ell- h_{\mu\nu}\bar{\Lambda} +\eta_{\mu\nu}\bar{\Lambda}_{\alpha\beta}h^{\alpha\beta},
    \end{split}
\end{equation}
with the parentheses representing the symmetrization of tensor indexes, and where we have defined
\begin{equation}\begin{split}
&\bar\Lambda_{\mu\nu} = \bar\lambda_{(\mu,\nu)} - \bar\lambda_\mu \bar\lambda_\nu, \\
& {\ell}_{\mu\nu}=  l_{(\mu,\nu)}-2\bar\lambda_{(\mu}l_{\nu)}-\bar\lambda^{\rho}
 h_{\rho(\nu, \mu)}+\frac{1}{2}\bar\lambda^{\rho}h_{\mu\nu, \rho} , \label{eq:ell}
 \end{split}
\end{equation}
with $\ell\equiv \ell_\mu^\mu$ and $\bar\Lambda\equiv\bar \Lambda_\mu^\mu$.

The energy-momentum tensor is also perturbed and may be split as
\begin{equation}
    T_{\mu\nu}= \bar T_{\mu\nu}+ \tau_{\mu\nu}, 
\end{equation}
where $\bar T_{\mu\nu}$ is the background energy-momentum tensor and all $ \tau_{\mu\nu}$ are small quantities when compared to $\bar T_{\mu\nu}$ for all $\mu,\, \nu$.

Taking into account the last approximations, at zeroth order in flat space-time, Eq.~\eqref{EE} implies in
\begin{equation}  
  \bar K _{\mu\nu}=8\pi G\,\bar T_{\mu\nu}. 
 \label{eq:EE0order}
\end{equation}
Relation \eqref{eq:EE0order} defines the background energy-momentum tensor in flat spacetimes to be, in general, different from zero. In the present theory, the flat Minkowski spacetime is fulfilled by a non-isotropic energy-momentum tensor given by $\bar T_{\mu\nu} = \bar\Lambda_{\mu\nu}-\eta_{\mu\nu} \bar\Lambda$. The important point here is that, since this energy-momentum tensor does not affect the geometry and, then, the trajectory of geodesic particles are straight lines, it cannot be detected by local experiments.  
On the other hand, the choice $\bar T_{\mu\nu}= 0$ requires the background vector $\bar{\lambda}_\mu$ must satisfy the condition $\bar \lambda_{(\mu,\nu)}-\bar \lambda_{\mu} \bar \lambda_{\nu}=0$. In this case, the solution for the background vector $\bar \lambda_{\mu}$ is given by $ 
    \bar \lambda_{\mu} =b_0\partial_\mu\ln\left(t+x+y+z\right)$,
where $b_0$ is a constant. 

Now proceeding with the linearzation of the field equations, it is well known that, at first order in $h_{\mu\nu}$ and its derivatives, the Einstein tensor reads
\begin{equation} \label{eq:linEinstein}
R_{\mu\nu}-\dfrac{1}{2}\eta_{\mu\nu}R=-\dfrac{1}{2}\square \gamma_{\mu\nu}- \dfrac{\eta_{\mu\nu}}{2} \gamma^{\sigma\rho}_{\;\;,\rho\sigma}+ \gamma^\sigma_{\;(\nu,\mu)\sigma},
\end{equation}
where $\gamma_{\mu\nu}=h_{\mu\nu}-\frac{1}{2}\eta_{\mu\nu}h$.

After replacing expressions \eqref{eq:linEinstein} and \eqref{eq:K01}  into Eq.~\eqref{EE}, it follows
\begin{equation}
  \begin{split}
  -\dfrac{1}{2}\square \gamma_{\mu\nu}- \dfrac{\eta_{\mu\nu}}{2} \gamma^{\sigma\rho}_{\;\;,\rho\sigma}+ \gamma^\sigma_{\;(\nu,\mu)\sigma}+ k_{\mu\nu}=8\pi G\, \tau_{\mu\nu},
 \label{V51}
    \end{split}
\end{equation}
where Eq.~\eqref{eq:EE0order} has been used and $G$ is taken as a constant parameter. 
In terms of $\gamma_{\mu\nu}$ the perturbation tensor $k_{\mu\nu}$ reads
\begin{equation}\label{eq:pertK}
  \begin{split}
k_{\mu\nu}=&\; l_{(\mu,\nu)} -2\bar\lambda_{(\mu}l_{\nu)} -\bar\lambda^{\rho}  \gamma_{\rho(\nu, \mu)} +\frac{1}{2}\bar\lambda^{\rho}\gamma_{\mu\nu, \rho} \\
 &\; -\eta_{\mu\nu}\left(l^\sigma_{\;,\sigma} -2\bar\lambda_\sigma l^\sigma  -\bar\lambda^\rho\gamma^\sigma_{\;\rho,\sigma}\right) \\
 &\;+\frac{1}{2}\bar\lambda_{(\mu}\gamma_{,\nu)} +\eta_{\mu\nu}\bar{\Lambda}_{\alpha\beta}\gamma^{\alpha\beta}-\gamma_{\mu\nu}\bar{\Lambda}.
     \end{split}
\end{equation}

In order to determine the physical properties of the metric perturbations in the present theory, we proceed as usual and consider the infinitesimal diffeomorphism generated by a vector field $\xi^\mu$, which gives rise to the coordinate transformation $x'^\mu= x^{\mu} + \xi^\mu(x)$. 
Taking notice that $\xi^\mu$ is an infinitesimal generator, it follows that the metric perturbations $h_{\mu\nu}$, the perturbation of the Herglotz vector $l_\mu$, and the energy-momentum tensor $\tau_{\mu\nu}$, transform respectively as
\begin{equation}
\begin{split}
     h^\prime_{\mu\nu}&=h_{\mu\nu}-\xi_{\mu,\nu}-\xi_{\nu,\mu},\\
   l^\prime_\mu&=l_\mu +\xi^\rho\bar\lambda_{\mu,\rho}-\bar\lambda_\rho\xi^\rho_{\;,\mu},\\
   \tau^\prime_{\mu\nu}&=\tau_{\mu\nu}+\xi^\rho \bar T_{\mu\nu,\rho}- \xi^\rho_{\;,\mu} \bar T_{\rho\nu} -\xi^\rho_{\;,\nu} \bar T_{\mu\rho} .
   \label{V53}
\end{split}
\end{equation}

From Eqs.~\eqref{V53} and \eqref{eq:K01b} we obtain the variation of the vector $k_{\mu\nu}$ in the form
\begin{equation}\label{eq:primek}
    k'_{\mu\nu}= k_{\mu\nu}+ \xi^\rho\bar K_{\mu\nu,\rho} -\xi^\rho_{\;,\nu}\bar K_{\mu\rho}
    -\xi^\rho_{\;,\mu}\bar K _{\nu\rho}.
\end{equation}
Now, by using Eqs.~\eqref{eq:EE0order} and \eqref{eq:primek} it follows
\begin{equation}\label{eq:primekb}
    k'_{\mu\nu}= k_{\mu\nu}+ 8\pi G\,\big( \xi^\rho\bar T_{\mu\nu,\rho} -  \xi^\rho_{\;,\nu}\bar T_{\mu\rho} - \xi^\rho_{\;,\mu}\bar T _{\nu\rho}\big),
\end{equation}
where we assumed a constant $G$.

Hence, given that the source on the right-hand side of Eq.~\eqref{V51} transforms according to the last relation in \eqref{V53}, i.e., $\delta\tau_{\mu\nu} = \xi^\rho\bar T_{\mu\nu,\rho}-
  \xi^\rho_{\;,\nu}\bar T_{\mu\rho} -\xi^\rho_{\;,\mu}\bar T_{\nu\rho}$, the comparison between this and Eq.~\eqref{eq:primekb} shows that the first order perturbation equations are gauge invariant.

Now due to the diffeomorphism invariance in a background spacetime region where Eq.~\eqref{eq:EE0order} is obeyed, we are free to make a gauge choice. In the present case, we choose the modified gauge condition 
\begin{equation}
    \gamma_{\mu,\rho}^{\;\rho}-\bar\lambda^\rho\gamma_{\mu\rho}+\frac{1}{2}\Bar{\lambda}_\mu \gamma+l_\mu=0,\label{eq:V66}
\end{equation}
 to simplify the field equations \eqref{V51}. With such a choice, the perturbation equations are cast as
 \begin{equation}
  \begin{split}
      &\;\square\gamma_{\mu\nu}-  \bar\lambda^{\rho}\gamma_{\mu\nu, \rho}+2\bar{\Lambda}\gamma_{\mu\nu}-
      2\gamma_{\rho(\mu}\bar\lambda^\rho_{\;, \nu)}+\gamma\bar\lambda_{(\mu, \nu)}\\
 &-\eta_{\mu\nu}\left(\bar\lambda^\rho\gamma_{\rho\;\;,\sigma}^{\;\sigma}+\frac{1}{2}\bar\lambda^\sigma\gamma_{,\sigma}+2\bar{\Lambda}^{\alpha\beta}\gamma_{\alpha\beta}-l^\sigma_{\;,\sigma}+4\bar\lambda_\sigma l^\sigma \right)\\ 
 &+ 4\bar\lambda_{(\mu}l_{\nu)}= -16\pi\, G\, \tau_{\mu\nu}.\label{V65}
    \end{split}
\end{equation}

Since $l_\mu$ is not a dynamic field, it does not propagate through space-time and then all degrees of freedom associated with it may be eliminated. Hence, from now on we assume $l_\mu=0$. Additionally,  conditions \eqref{eq:V66} imply that four metric degrees of freedom are fixed. This means that the metric tensor still has six degrees of freedom, some of them may not be physical, or may not propagate, and a further detailed analysis is necessary. This is an important study that we prefer not to present here to avoid a too lengthy text.

\subsection{Plane wave decomposition of the gravitational perturbations}

For simplicity, we now assume that $\bar\lambda_\mu$ is a constant vector and that the perturbed energy-momentum tensor is zero, $\tau_{\mu\nu}$=0.
After that, and by taking $l_\mu=0$ into Eqs.~\eqref{eq:V66} and \eqref{V65} it follows
\begin{equation}
  \begin{split}
 &  \gamma_{\mu,\rho}^{\;\rho}-\bar\lambda^\rho\gamma_{\mu\rho}+\frac{1}{2}\Bar{\lambda}_\mu \gamma= 0,\\
&  \square\gamma_{\mu\nu}-  \bar\lambda^{\rho}\gamma_{\mu\nu, \rho}+2\bar{\Lambda}\gamma_{\mu\nu} \\
     &  \;-\eta_{\mu\nu}\left(\bar\lambda^\rho\gamma_{\rho\;\;,\sigma}^{\;\sigma}+\frac{1}{2}\bar\lambda^\sigma\gamma_{,\sigma}+2\bar{\Lambda}^{\alpha\beta}\gamma_{\alpha\beta}\right)=0 . \label{eq:V65new}
    \end{split}
\end{equation}
We then look for solutions to the last equations in the plane wave form
\begin{equation}
\begin{split}
    \gamma_{\mu\nu}=A_{\mu\nu}e^{ik_\sigma x^\sigma},\label{V66}
    \end{split}
\end{equation}
where $A_{\mu\nu}$ is a constant and symmetric tensor, and $k_\mu$ is the wave vector. By taking the expression \eqref{V66} into \eqref{eq:V65new} it follows 
\begin{equation}
   \begin{aligned}
    &A\bar\lambda_\mu +2\left(ik^\nu-\bar\lambda^\nu \right)A_{\mu\nu}=0,\label{eq:V78}\\
       & k^2+i\bar\lambda_\mu k^\mu+2\bar\lambda^2=0,\\
        & \frac{i}{2}A\bar\lambda_\mu k^\mu+A_{\mu\nu}\bar\lambda^\nu\left(ik^\mu-2\bar\lambda^\mu\right)=0. 
    \end{aligned} 
\end{equation}

The admission of a plane wave solution for a constant Herglotz vector field $\lambda_\mu$ imposes five restrictions to the amplitude tensor $A_{\mu\nu}$, namely, the first (four) and the last (one) relations in \eqref{eq:V78}. Hence, five components are left undetermined, and deeper analysis to give the physical interpretation of them is necessary. As commented above, this study is beyond the scope of the preset work. 

Now, the constancy restriction on $\bar\lambda$ would pick out a preferential direction in spacetime, since this vector couples to the metric perturbation tensor. In other words, the local Lorentz symmetry is broken and the wave propagation may not be isotropic even in flat spacetime. Additionally, the second relation in \eqref{eq:V78} that gives a complex dispersion relation translates into damped gravitational perturbations, as we shall see below.

\subsection{Damping gravitational perturbations} 

Since the Herglotz vector is arbitrary, we may make further assumptions to simplify the analysis.  Here we assume that  $\bar\lambda_\mu$ is a light-light vector, i.e., $\bar\lambda^2=0$.  

As it is well known, the complex form of the dispersion relation, as the second relation in \eqref{eq:V78}, leads to dissipative effects on the wave propagation. Indeed, the components of the wave vector $k^\mu$ assume complex values and the imaginary parts contribute to the damping or forcing of the wave amplitude, depending on the vector $\bar\lambda_\mu$.
To explore this dependence, we split the timelike and spacelike components of vectors $\bar\lambda_\mu$ and $k_\mu$, respectively, as $\bar\lambda_\mu=(-\bar\lambda_t,\, \bm{\bar\lambda})$ and $k_\mu = (-\omega,\, \bm{k})$. Therefore, from the second equation in \eqref{eq:V78} one finds the dispersion relation  
\begin{equation} \label{eq:dispersion}
    \omega=\frac{1}{2}\left[\alpha_+ - i\left(\pm \alpha_- +\bar\lambda_t\right)\right],
\end{equation}
with $\alpha_\pm$ defined by
\begin{equation}\label{eq:alphas}
     \begin{split}
    \alpha_\pm = & \bigg[
    \sqrt{\left(2|\bm{k}|^2-\bar\lambda_t^2/2\right)^2+\left(2\bm{k}\cdot \bm{\bar\lambda}\right)^2}\\
    & \pm \left(2|\bm{k}|^2 -\bar\lambda_t^2/2\right) \bigg]^{1/2},
    \end{split}
\end{equation} 
where the dot ($\cdot$) stands for the scalar product, and we have chosen the solution for which the real part of $\omega$ is non-negative.
 After \eqref{eq:dispersion}, the exponential part of the solution in \eqref{V66} goes as $\exp[\left(\pm\alpha_- +\lambda_t\right)t/2\big]$ times an oscillatory function of time. This indicates that the wave may be damped (or amplified) while traveling throughout spacetime. 

The effects of the Herglotz field $\lambda_\mu$ on the wave propagation are more easily identified in two particular cases, namely, the case where the wave vector $\bm{k}$ is orthogonal to $\bm{\bar\lambda}$ and the case where these two spacial vectors are parallel to each other. 

Taking the case where $\bm{k}$ is parallel to $\bm{\bar\lambda}$ it follows that $\alpha_+ =2|\bm{k}|$, $\alpha_-=\lambda_t$, and  $\omega =|\bm{k}| -i\lambda_t$. This result follows by noticing that, in the present situation, without loss of generality, we may choose coordinate axes so that the wave and Herglotz vectors take the forms $k_\mu=\left(-\omega,\, 0,\, 0,\, k_3\right)$ and $\bar\lambda_\mu=\left(-\bar\lambda_t,\, 0,\, 0,\, \pm\bar\lambda_t\right)$, with $k_3=\pm |\bm{k}|$ and where the assumption $\bar\lambda^2=0$ was used to write $\bar\lambda_3=\pm\bar\lambda_t$. Hence, one has $\bm{k\cdot\bar\lambda} = \pm k_3 \bar\lambda_t$ and from Eqs.~\eqref{eq:dispersion} and \eqref{eq:alphas} the just stated result follows. As a consequence, the wave amplitude depends on an exponential function of time that depends on the Herglotz parameter $\bar\lambda_t$, namely, the exponential factor is exactly $\exp\left[\lambda_t t\right]$ and  so the wave is damped in the case $\lambda_t <0$ and it is amplified in the case $\lambda_t> 0$.

Now choosing the particular case with $\bm{k\cdot\bar\lambda}=0$ two situations come out. The first solution is $\alpha_+ =2\sqrt{|\bm{k}|^2-\lambda_t^2/2}$, $\alpha_-=0$, for $|\bm{k}|^2-\lambda_t^2/2>0$. In this case it follows $\omega =\left(\alpha_+ -i\lambda_t\right)/2$, what implies the wave amplitude varies with time as $\exp[\lambda_t t/2\big]$. Therefore, as in the preceding case, the wave is damped in the case $\lambda_t <0$ and it is amplified in the case $\lambda_t> 0$. 
The second situation is for  $|\bm{k}|^2-\lambda_t^2/2<0$, which gives $\alpha_+ =0 $ and $\alpha_-=\sqrt{\lambda_t^2/2 -|\bm{k}|^2}$. In this case it follows $\omega =-i \left(\pm \alpha_- + \lambda_t\right)/2$, what implies the wave amplitude varies with time as $\exp[\left( \pm\alpha_-+\lambda_t\right) t/2\big]$, and since one has $0\leq \alpha_-\leq|\lambda_t|$, again the wave is damped in the case $\lambda_t <0$ and it is amplified in the case $\lambda_t> 0$.

\subsection{The wave speed}

The speed of the perturbation waves may be determined from the above results. We start by studying the phase speed $v_f$, defined by $v_f=\big|\Re(k_0)\big|/\big|\mathbf{k}\big| = \omega/k$. Using relations \eqref{eq:dispersion} and \eqref{eq:alphas} we see that the phase speed depends on the wavelength, on the propagation direction, and on the strength of the Herglotz $\lambda_\mu$. The dependence of the phase speed on the propagation direction is seen more clearly by taking two particular cases for which the relation turns out simple.

The first case with $\bm{k}$ is parallel to $\bm{\bar\lambda}$, where one has and  $\omega =|\bm{k}| -i\lambda_t$. In this case the phase speed of the wave is given by
\begin{equation}
    v_f= \dfrac{ \Re(\omega )}{|\mathbf{k}|}=1, 
    \label{eq:wavespeed}
\end{equation}
which is exactly the speed of light.    

A second simple case is when the propagation is orthogonal to the spacelike Herglotz vector, i.e., for $\bm{k\cdot\bar\lambda}=0$. 
Here, it follows  
\begin{equation}
    v_f= \sqrt{1 - \frac{\lambda_t^2}{2|\bm{k}|^2}}.
    \label{eq:wavespeed2}
\end{equation} 
This result seems to imply that the Herglotz parameter $\lambda_t$ imposes a cutoff for the propagation of plane waves, no propagation for low wavenumber values compared to $\lambda_t$. For large wavenumbers, the phase speed approaches the speed of light. 
However, one must also consider the group velocity and, more precisely, the speed of propagation of energy and momentum in the present theory. A simple calculation by using the definition $v_g= \partial \Re(\omega)/\partial k $ furnishes the group speed, and the resulting expressions show that the group velocity may be larger than the speed of light. However, a deeper analysis is necessary to investigate whether the energy transported by gravitational waves really may travel faster than the speed of light, but this analysis is beyond the goals of the present work.
 
\vspace{-.25cm}

\subsection{Estimating the Herglotz parameter}

Comparing the linear regime of the theory proposed here and the linear regime of that one proposal in \cite{Lazo:2017udy}, we see that there is a difference between them. The difference is in the speed of the wave propagation. 
Here the speed of wave propagation coincides with the speed of light, although the same is possible in theory at \cite{Lazo:2017udy} by choosing the same coupling vector in the form $\bar\lambda_\mu=\left(-\bar\lambda_t,0,0,\pm\bar\lambda_t\right)$. However, the proposal made ensures that the wave always propagates at a speed slower than the speed of light.

The recent data from gravitational waves detection may be used to estimate the parameter $\lambda_T$. Taking for instance the event GW170817 \cite{TheLIGO:2017qsa}, which is located at the distance of about 130 million light-years from Earth, and using the result \eqref{eq:wavespeed} we find the time-travel $\tau$ of waves produced at that event is of about 130 million years. Let $R$ be the relation between the supposedly damped wave (observed in the Earth) and the amplitude of the corresponding nondamped wave. The theoretical prediction for $R$ is obtained from the analysis of the last section, that is $R=\exp\left[\lambda_t\tau\right]$, or equivalently $\lambda_t\tau \simeq \ln R$. Assuming further that the amplitude damping due to dissipative effects is of the order of the initial amplitude we have $R=1/2$ and it follows $|\lambda_t|\sim \times 10^{-16}{\rm s}^{-1}$. This is to be considered an upper bound for the parameter $\lambda_t$, since the amplitude damping is bounded by the statistical error bars on the observed amplitude of the wave.  
The more distant the gravitational source is, the sharper is the upper bound on $\lambda_t$. In fact, using the event of Ref.~\cite{Abbott:2016blz} which is estimated to have occurred at the redshift of the order of $z=0.1$, which means that the wave has traveled about $1.3$ billion years to reach Earth, we find $|\lambda_t|\sim 10^{-17}{\rm s}^{-1}$, yet one order of magnitude larger than what is estimated from cosmology (see Sec.~\ref{sec:inflat}).

 \section{Final remarks}
\label{sec:conclusion}

We have considered the Herglotz variational principle for fields to propose a totally covariant nonconservative gravity formulation. As a result, by using the usual gravity Lagrangian density and introducing an arbitrary background vector field (the Herglotz field), we obtained the modified gravitational field equations that present a totally tensorial structure. Therefore, the non-tensorial character of the theory obtained in \cite{Lazo:2017udy} is solved.  

When the theory was put to the test, within the scope of cosmology using the FLRW geometry, different types of solutions were obtained assuming the conservation and nonconservation of the energy-momentum tensor.  In the case of conserved energy-momentum tensors, it results in solutions for the scale factor such as power-law and exponential forms, the last form being appropriate for both inflation with big bang and a late phase of accelerated expansion. 
In this case, the extra (Herglotz) vector field, which in FRLW spacetimes has only one nontrivial component $\phi$, plays a role similar to the cosmological constant in general relativity, and an estimate for its numerical value at present time was obtained by considering the present value of the Hubble parameter.  
In the case of non-conserved energy-moment tensors, we have found the inflation standard solution among other interesting solutions. Among these, the solution obtained in subsection \ref{sec:CDM} stands out, which represents a universe filled with dust (cold dark matter), over a period of time, thus avoiding the introduction of a dark energy component (or a dilaton field) to explain the accelerated expansion (or the inflationary) period.

Despite a variety of types of cosmological solutions that can be found in view of the arbitrariness of the Herglotz field $\phi$, the simplest cosmological solutions have some issues to considered and further investigated. One of them is the existence of solutions with negative energy density, leading us to restrict the choices for $\phi$ that result in non-negative energy density. 
Another issue is the same as it happens in general relativity. Due to the restriction imposed by choosing an equation of state in the form $p=\omega \rho$, with $\omega$ being a constant parameter, the theory is unable to provide an accelerated expansion in late cosmological times without the aid of an exotic material known as dark energy. However, in the present theory, these apparent flaws may be remedied by introducing some kind of mechanism to select the appropriate $\phi(t)$ for each phase of the universal expansion.

The linear regime of the theory was also studied and wave-like solutions were shown to exist. As expected, this study revealed the dissipative behavior of gravitational wave propagation, which can be forced or dampened depending on the Herglotz four-vector $\lambda_\mu$. In the present theory, the gravitational waves propagate with the speed of light and, just like in general relativity, present two modes of propagation. 
 Again, considering a plane wave propagating along a specific direction, a numerical estimate for the extra (Herglotz) parameter was obtained by using the recent data on gravitational waves. 
 
 The applications presented in the present work should be considered as a preliminary analysis, so that further and deeper studies are necessary to test the theory against observational data. Our immediate interest is to investigate the existence of solutions representing compact objects in this nonconservative gravity theory.

\section*{Acknowledgments}
J. A. P. P. was financed in part by Coordenação de Aperfeiçoamento
de Pessoal de Nível Superior (CAPES), Brazil, Finance Code 001.
M. J. Lazo thanks Conselho Nacional de Desenvolvimento Científico e Tecnológico (CNPq), Brazil, Grant No. 310386/2020-9 and Grant No. 425333/2018-3.
V. T. Z. thanks CAPES, Brazil, Grant No. 88887.310351/2018-00, and CNPq, Brazil, Grant No. 309609/2018-6.

\end{document}